\documentclass[
 reprint, prd, 
 amsmath,amssymb,
 aps,nofootinbib
]{revtex4-2}

\newcommand{\dam}[1]{{\bf \color{red} #1}}

\usepackage[breaklinks=true,colorlinks=true,citecolor=blue,urlcolor=blue,linkcolor=blue,pdfencoding=auto,psdextra]{hyperref}
\usepackage{times}
\usepackage{epsfig}
\usepackage{graphicx}
\usepackage{epstopdf}
\usepackage{bm}
\usepackage{color}
\usepackage[english]{babel}
\usepackage{microtype}

\usepackage{xcolor}

\newcommand{\s}{\,s$^{-1}$}
\newcommand{\ergs}{\mbox{\,erg\,\s}}

\renewcommand{\deg}{^{\circ}}

\newcommand{\rj}{R_{\rm jet}}
\newcommand{\rc}{R_{\rm cenA}}

\newcommand{\qi}{q}

\def\app#1#2{%
  \mathrel{%
    \setbox0=\hbox{$#1\sim$}%
    \setbox2=\hbox{%
      \rlap{\hbox{$#1\propto$}}%
      \lower1.1\ht0\box0%
    }%
    \raise0.25\ht2\box2%
  }%
}

\newcommand{\aap}{A\&A}

\newcommand{\apjl}{ApjL}
\newcommand{\apjs}{ApjS}

\begin{document}
\title{Ultrahigh-energy neutrinos as a probe of espresso-shear acceleration in jets of Centaurus A}

%\author[0000-0001-9475-5292]{Rostom Mbarek}
\author{Rostom Mbarek}
\altaffiliation{Neil Gehrels Fellow}
\email{rmbarek@umd.edu}
\affiliation{Joint Space-Science Institute, University of Maryland, College Park, MD, USA}%
\affiliation{Department of Astronomy, University of Maryland, College Park, MD, USA}
\affiliation{Astrophysics Science Division, NASA Goddard Space Flight Center, Greenbelt, MD, USA}

%\author[0000-0003-0939-8775]{Damiano Caprioli}
\author{Damiano Caprioli}
\affiliation{University of Chicago, Department of Astronomy \& Astrophysics, 5640 S Ellis Ave., Chicago, IL 60637, USA}
\affiliation{Enrico Fermi Institute, The University of Chicago, Chicago, IL 60637, USA}

%\author[0000-0003-0939-8775]{Kohta Murase}
\author{Kohta Murase}
\affiliation{Department of Physics, The Pennsylvania State University, University Park, Pennsylvania 16802, USA}
\affiliation{Department of Astronomy \& Astrophysics, The Pennsylvania State University, University Park, Pennsylvania 16802, USA}
\affiliation{Center for Multimessenger Astrophysics, Institute for Gravitation and the Cosmos, The Pennsylvania State University, University Park, Pennsylvania 16802, USA}
\affiliation{Center for Gravitational Physics and Quantum Information, Yukawa Institute for Theoretical Physics, Kyoto, Kyoto 606-8502 Japan}

\begin{abstract}
It has been suggested that Centaurus A (Cen A) could make a contribution to the observed ultrahigh-energy cosmic-ray (UHECR) flux. 
We calculate the flux of astrophysical neutrinos produced by UHECRs accelerated in the jet of Cen A, a close-by jetted active galactic nucleus. We use a bottom-up approach, in which we follow the energization of protons and heavier elements in a magnetohydrodynamic simulation of a relativistic jet with proper parameters of Cen A, also accounting for attenuation losses based on the observed photon fields. 
We compare the expected neutrino flux with the sensitivity of current and planned neutrino detectors. We find that the detection of $\sim 10^{17}$--$10^{18}$~eV neutrinos from Cen A would require ultimate neutrino detectors that reach a point source sensitivity of $\sim {\rm a~few}\times10^{-13}~{\rm erg}~{\rm cm}^{-2}~{\rm s}^{-1}$. Successful detection, though challenging, would be useful in constraining the Cen A contribution to the UHECRs. 
\end{abstract}

\defcitealias{murase+14}{MID14}
\defcitealias{mbarek+19}{MC19}
\defcitealias{mbarek+21a}{MC21}
\defcitealias{mbarek+23}{MCM23}

\date{\today}
\maketitle

\section{Introduction}

Jets in active galactic nuclei (AGNs) are promising sources of ultrahigh-energy cosmic rays (UHECRs) and neutrinos up to $\gtrsim 10^{18}$eV. 
These jetted AGNs satisfy the confinement requirements for UHECRs, e.g., \citep[][]{cavallo78,hillas84}, have luminosities large enough to sustain their energy injection rate, e.g., \citep[][]{katz+09,murase+19,jiang+21}, and allow for a heavy elemental composition at the highest energies, e.g., \citep[][]{Mur+12,mbarek+23}, being consistent with observations by the Pierre Auger Observatory \citep{auger14b,auger17,yushkov+19}. 
Recent observations by the Telescope Array (TA) have also supported that UHECRs are dominated by heavy nuclei at the highest energies \citep{TA24}.

Centaurus A (Cen~A) is a close-by radio-loud FR I\footnote{Fanaroff-Riley (FR) I are radio-loud galaxies with extended jets that are typically decelerated to nonrelativistic bulk flows within $\sim1$~kpc \citep[e.g.,][]{wardle+97,arshakian+04,mullin+09}.} AGN exhibiting a jet and has been extensively considered in the literature as a possible source of UHECRs, e.g., Refs.~\citep[][]{honda+09,biermann+12,wykes+13,murase+14}, especially after the Auger experiment has shown a significant flux excess in its direction \citep{auger18b,AUGER22,AUGER23ICRC}, with a best-fit contribution of $\gtrsim 3-25\%$ at an energy $E\simeq 40$~EeV and $\simeq 10-50\%$ at $E\simeq 100$~EeV, depending on the redshift evolution \citep{AUGER24}. 
Cen A could even make the dominant contribution to the UHECR flux above the ankle energy~\cite{kim13,Eichmann:2017iyr,deOli+20,mollerach+24}.   

This makes this source particularly interesting for ultrahigh-energy (UHE) neutrinos, which have been discussed for this source, and deemed to likely originate from the core~\citep{cuoco+08,Sah+12,Fra+16}, large-scale jets~\citep{Kac+09,Art+12,zhang+23}, or propagation~\citep{Der+09,deOli+20}.

In \citet{mbarek+19} and \citet{mbarek+21a}, we have shown that galactic cosmic-ray (CR) seeds can be reaccelerated to UHECR levels in transrelativistic or mildly relativistic radio-loud AGN jets via the so-called \emph{espresso} mechanism \citep{caprioli15} and shear acceleration \citep{kimura+18};
the resulting chemical composition, spectrum, and anisotropy features can be consistent with UHECR phenomenology.
We have also analyzed the effects of attenuation losses due to different photon fields in \citet{mbarek+23} to present a global scheme that accounts for particle injection, particle acceleration, spectra of UHECRs with energies above $10^{18}$eV, and the ensuing neutrino spectrum from radio-loud AGN jets. 
The jet of Cen~A is expected to have an intrinsic velocity of $\sim0.6$~c at kiloparsecs~\citep{Wyk+19}, and it is possible to expect \emph{espresso} acceleration at smaller scales in the jet spine and shear acceleration mechanism at larger scales.

In this study, we go beyond previous efforts that calculated the Cen~A UHE neutrino flux, by accounting for the properties of particle acceleration and propagation in state-of-the-art jet simulations and recent observation-based modeling of the photon fields in Cen~A.
In a nutshell, we employ the methods introduced in \citet{mbarek+23} to estimate the neutrino flux expected from Cen~A's jet over a substantial energy range. 
We eventually compare our results with the resolution and sensitivity of current and future neutrino observatories, and set limits on the elusive highest-energy neutrinos with $E_\nu > 10^{17}$eV from such a source, e.g., Refs.~\citep[][]{ARA16-long,Arianna17-long,ICECUBE-GEN2-14,olinto+17,GRAND20,PUEO21}.

\section{MHD Jet Setup}\label{sec:phys-param}
We propagate ions in a relativistic magnetohydrodynamic (MHD) simulation of an AGN jet with properties similar to those expected of Cen~A (size, structure, and magnetic field), augmented with different sub-grid scattering (SGS) prescriptions \citep{mbarek+21a} and the multi-wavelength photon fields expected in the Cen~A environment\footnote{Note that the photon fields are ascribed based on observations, and our MHD simulations do not have a radiative component.} (see section~\ref{sec:cenA-fields}). 
We then calculate the expected neutrino spectrum produced as a result of UHECR acceleration in the jet, taking into account different production channels, including photomeson production interactions and neutrinos from the $\beta$ decay of neutrons produced through nuclear photodisintegration \citep{mbarek+23}. Proton-proton interactions should be negligible \citep{mbarek+23}.
%Initially, we calculate the expected neutrino flux from CenA as a point source based on the methods presented in \citet{mbarek+23} over an extended energy range. 

\subsection{Jet Initialization}
As in \citet{mbarek+23}, we consider a three-dimensional relativistic MHD simulation performed with the {\tt PLUTO} code \citep{mignone+10}. 
The box size measured 48~$\rj$ in the $x$- and $y$-directions and 100~$\rj$ in the $z$-direction, where $\rj$ is the jet radius, in a grid that has $512 \times 512 \times 1024$ cells with four adaptive mesh refinement levels (see \citep{rossi+08} for more details about the computational setup). $\rj$ is defined as the magnetization radius of the jet, which defines the scale around which the magnetic field forces become dominant and significantly influence the dynamics of the conducting fluid. A more thorough description is included in, e.g., Fig.~4 in \citep{mbarek+19}.
The jet/ambient density contrast is set to $10^{-3}$, the jet sonic number to 3 and Alfv\'enic Mach number to $1.67$.
After the jet has developed, the effective Lorentz factor $\Gamma_{\rm eff}$ in the inner regions of the jet sits at $\Gamma_{\rm eff}\sim 3.5$. 
Although a velocity of $0.2-0.7$~c has been associated with Cen~A's jet \citep{Wyk+19}, this $\Gamma_{\rm eff}$ value should not be surprising as the inner spinal regions of the jet are expected to have large Lorentz factors, e.g., \citep[][]{chiaberge+01,ghisellini+05}, with no appreciable differences between FR-I and FR-II jets in their initial $\Gamma_{\rm eff}$ values \citep{giovannini+01,cg08}.

%\subsection{Jet Physical Parameters}
In a relativistic MHD simulation, we can associate physical energies to seed particles of charge $Z$ by setting the physical values of  jet radius $\rj$ and magnetic field $B_0$, since CR gyroradii are normalized to $\rj$ and $B_0$. 
Here, we fix the effective jet radius $\rj$ based on the observed extent of the jet of $\gtrsim 1$~kpc \citep{hardcastle+06}, such that $\rj = 15$~pc. 
The observed radial extent might be larger than $\rj$, but powerful jet spinal regions where acceleration occurs should be more compact, e.g., \citep[][]{mbarek+19}.
As for the magnetic field, \citet{goodger+10} inferred an equipartition value reaching 250-750$\mu$G in the jet knots, but we set  $B_0 = 100~\mu$G as the diffuse jet component should be lower.
Powerful FR-II jets are routinely inferred to have such large $B_0$, but the spines of FR-I jets could also reach these levels, e.g., \citep[][]{hardcastle+04,hawley+15}. 
These parameters, along with the kink that develops (see \citet{mbarek+23}), can mimic the physical conditions of Cen~A's jet.
However, our final considerations do not depend strongly on the uncertainties in these parameters.

\subsection{Jet photon fields}\label{sec:cenA-fields}

\begin{figure}
\centering
\includegraphics[width=0.48\textwidth,clip=false, trim= 0 0 0 0]{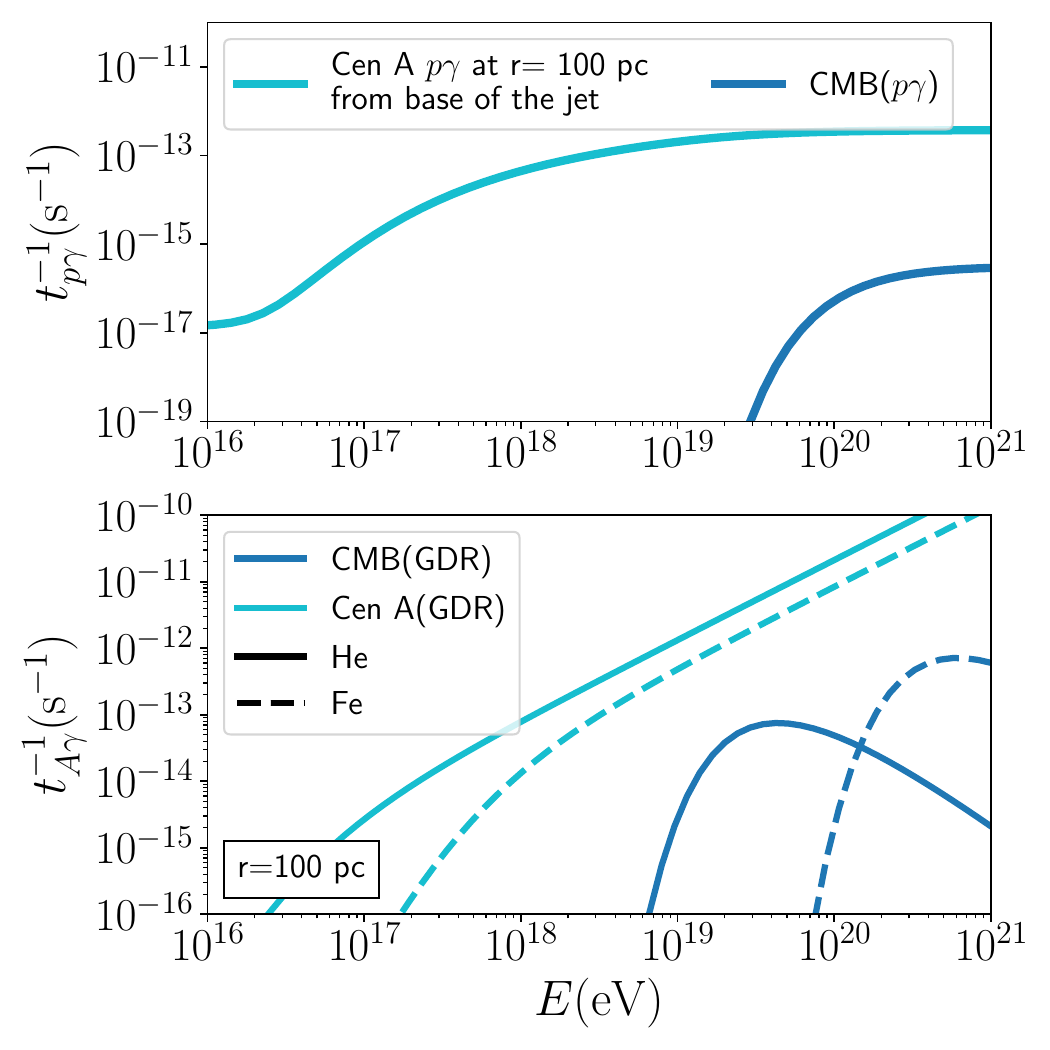}
\caption{Upper Panel: Photomeson production cooling time for a proton located at 100~pc away from the jet base along the jet axis for a radius $\rc = 15$~pc. 
Lower Panel: Same as the upper panel but for photodisintegration interactions of helium (He) and iron (Fe) nuclei, based on GDR cross section.}
\label{fig:ph-fields}
\end{figure}

While propagating in the jet, tracked particles experience both photomeson production ($p\gamma$) and photodisintegration ($A\gamma$) interactions, which produce different populations of high-energy neutrinos \citep{mbarek+23}.
We then augment our relativistic MHD jet structure with external radiation based on Cen~A's jet observations, where an extended multi-wavelength spectral energy distribution is modeled including the non-thermal inner-core emission, as discussed in \citet{zhang+23}, where the seed photon density is implemented as $L_{j}\delta^4/(4\pi r^2)$, such that $L_{j}$ is the jet luminosity, and $\delta$ the Doppler boosting.
This allows us to estimate the photon energy density at a distance $r$ from the jet base \citep[See Equation~21 in][for more details]{zhang+23}, and thus the local probability for $p\gamma$ and $A\gamma$ interactions \citep{mbarek+23}.
The upper panel of Figure~\ref{fig:ph-fields} shows the distance-dependent $p\gamma$ cooling time (see Equation~B5 in \citet{mbarek+23}), while the lower panel of Figure~\ref{fig:ph-fields} shows the $A\gamma$ cooling times for nuclei with energy $E_A$, assuming the giant dipole resonance (GDR) cross section (see Equations~C10 and~C12 in \citet{mbarek+23} and references therein).

\section{Neutrino yield from the Centaurus A Jet}\label{sec:cenA-flux}
\subsection{Neutrino flux from Cen~A's large-scale jet}
We consider two main channels for the production of source neutrinos from Cen~A's jet: \emph{i)} ($p\gamma$) interactions and \emph{ii)} the $\beta$ decay of secondary neutrons produced during the photodisintegration ($A\gamma$) of heavy nuclei.
Following \citet{mbarek+23} we write the all-flavor neutrino flux as:
\begin{equation}\label{eq:Numain}
	\begin{aligned}
		E_\nu^2 \phi_\nu (E_\nu) \approx  E^2 \phi_{\rm CenA}(E \approx 20A E_\nu)  f_\nu(E \approx 20A E_\nu, \qi),
	\end{aligned}
\end{equation}
where $E^2\phi_{\rm CenA}(E \approx 20A E_\nu)$ is the UHECR flux from Cen~A at $E$, $A$ is the atomic mass, and $f_\nu$ is the conversion factor from the UHECR flux to the neutrino flux, which depends on the injection slope $q \in [1,2]$ \citep{mbarek+23} of escaping UHECRs.
Here 
\begin{equation}
f_\nu (E)\approx \frac{3}{8} f_{\rm mes}(E),     
\end{equation}
where $f_{\rm mes}$ is the effective optical depth that is conventionally used in the literature~\cite{murase+10}, and $f_\nu$ accounts for the energy fraction carried by neutrinos that coming from the fact that the number ratio of charged pions to neutral pions is unity.   

%Such an energy is chosen as reference because it is these UHECRs that contribute most of the expected neutrinos \citep{mbarek+23}. \dam{Did I get this right?}

The conversion factor $f_\nu(E, \qi)$ is calculated directly by propagating particles in the relativistic MHD simulations and accounting for the neutrino production, as discussed above.
%Generally speaking, this parametrization roughly captures the details of a full kinetic approach.
The factor $f_\nu$ depends on the UHECR injection slope $q$, which is not fully constrained, but generally lies within $1\leq \qi \leq2$. 
For a lighter UHECR composition, $\qi \sim 2$ is preferred, e.g., \citep[][]{waxman95,katz+09,aloisio+11,jiang+21}, but more recent data favor a heavy UHECR composition \citep{AUGER22,TA24} and harder spectra with $1 \leq \qi \leq 1.6$, e.g., \citep[][]{aloisio+11,gaisser+13,aloisio+14,taylor+15,jiang+21}. Further details on the dependence on $q$ can be found in \citet{mbarek+23}.
Importantly, $f_\nu(E_\nu, \qi)$ preserves a heavy chemical composition \citep{mbarek+23}, as observed by Auger \citep{AUGER22}.

As for the UHECR flux from Cen~A, we consider Auger's anisotropy hints where the average flux above 40 EeV from a 25$^\circ$ top-hat region centered around the Centaurus system, sits at $ \simeq 1.59 \times 10^{-2}$km$^{-2}$~yr$^{-1}$~sr$^{-1}$ \citep{AUGER22}. UHECRs could experience heavy deflections during their propagation from the source because Cen~A lives in a cluster with magnetic fields that could reach $0.1\mu G$ \citep{vallee+02,taylor+23}. This would mean that the integrated flux in the direction of Cen~A is most likely a lower limit on the expected UHECR flux from Cen~A\footnote{A correction based on the galactic magnetic fields could be needed \citep[e.g.][]{jansson+12}, however the excess region might be large enough to make galactic deflections inconsequential to the main conclusions of this paper.}.
The flux at $10^{19}$~eV centered around Cen~A reads $4 \pi E^2 \phi_{\rm CenA}(E_{19}) \simeq 4\times 10^{-10}$GeV~cm$^{-2}$~s$^{-1}$ for a single source emission.
A more optimistic estimate for the Cen~A flux would be a factor of $E^2 \phi_{\rm UHECR}(10 EeV)/E^2 \phi_{\rm UHECR}(40 EeV) \simeq 3$ larger, if we assume that the Cen~A flux has the same spectral features as the overall UHECR flux $E^2\phi_{\rm UHECR}$.
These estimates result in a minimum UHECR power $L_{\rm UHECR} \simeq 3 \times 10^{39}$\ergs $\simeq 3\times 10^{-4} L_{\rm CenA}$, where $L_{\rm CenA}\sim 10^{43} \ergs$ is the Cen~A jet power \citep{croston+09}. The energetics requirement is not demanding, and is consistent with the idea that FR-I jets can be UHECR sources.

\begin{figure}
	\begin{centering}
		\includegraphics[width=0.48\textwidth,clip=false,trim= 0 0 0 0]{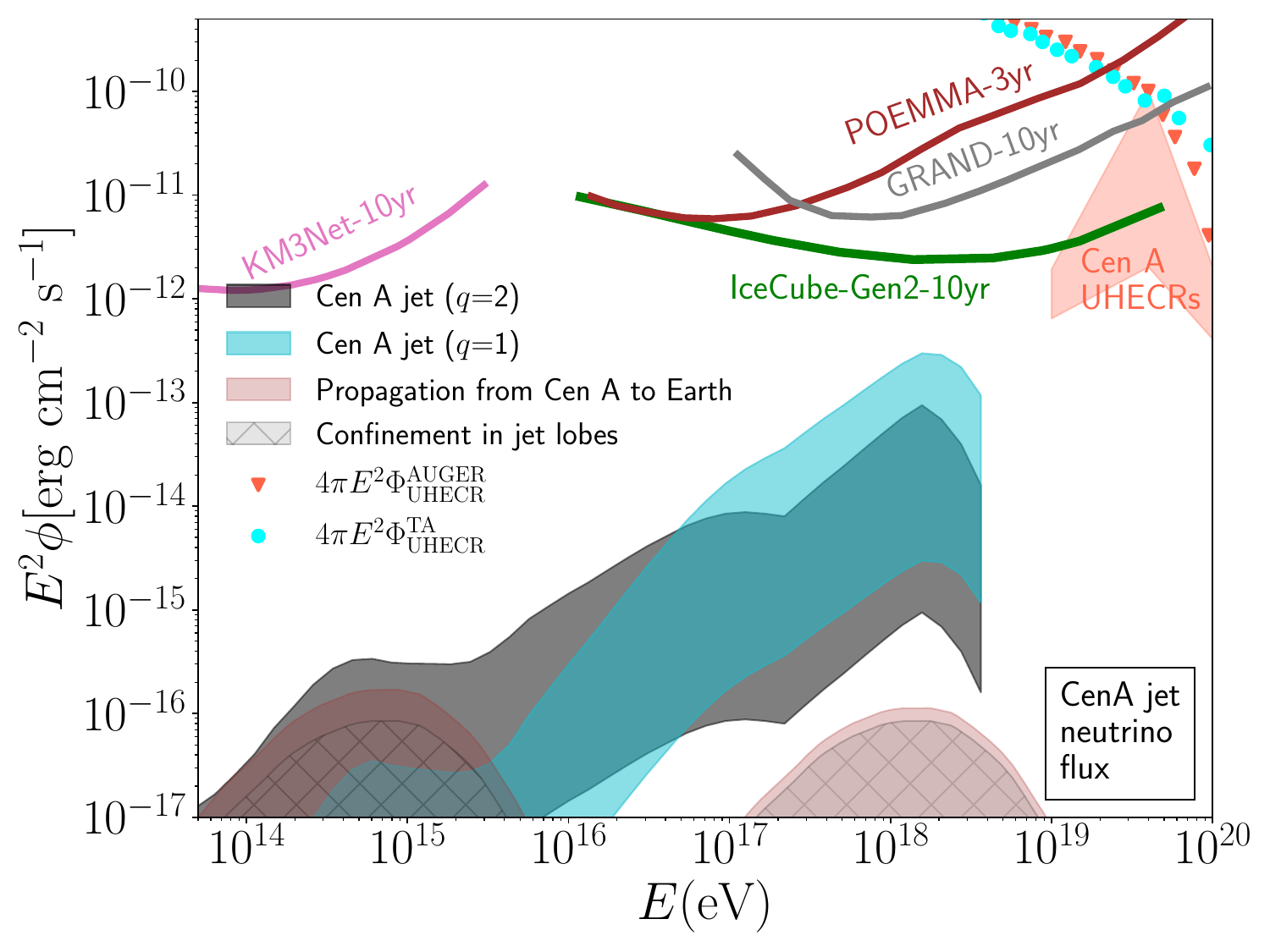}
		\caption{Expected neutrino flux from UHECRs accelerated in Cen~A's jet and propagated to Earth, along with the UHECR flux from Auger \citep{AUGER20} and TA \citep{TA16} with neutrino detection resolutions of GRAND \citep{GRAND20}, POEMMA \citep{POEMMA20}, KM3Net \citep{KM3NET24}, and IceCube-Gen2 radio \citep{ICECUBE-GEN2,ICECUBE-ICRC21} for comparison. Note that for GRAND and IceCube-Gen2 radio a declination of $\delta=-45$~deg is assumed.
        Cen~A's contribution to the UHECR flux at, e.g., 40~EeV is depicted between $\psi = 3\%$ and $\psi = 100\%$ \citep{AUGER24}. We include contributions from both propagation to Earth (in brown) and confinement within the lobes (hatched). These processes collectively yield two distinct signals: one at the EeV level, from photomeson interactions of UHECRs, and another at the PeV scale, resulting from the decay of neutrons generated during photodisintegration interactions.
        }
		\label{fig:cenA-flux}
	\end{centering}
\end{figure}

This top-hat anisotropic flux constitutes a fraction $\psi \sim 1\%$ of the overall UHECR flux at 40~EeV. More recent data modeling is more optimistic and fixes the best-fit flux in the direction of Cen~A at $\psi \sim 3-25\%$ at 40~EeV\footnote{Note that $\psi = 25\%$ is obtained in the case of flat evolution.} and  $\psi \sim 10-50\%$ for 100~EeV UHECRs \citep{AUGER24}.
%However, uncertainties in the signal are quite large and Cen~A's contribution can rise to 10-50\% at 100 EeV within 1$\sigma$\citep{AUGER24}. 
Previous models also find that a contribution of $\sim10$\% from Cen~A are possible considering the large-angle deflections during propagation \citep[e.g.][]{wykes+18}. Other studies also point out that hints of anisotropies detected by TA in the direction of M81 and M82 are echoes of UHECRs emitted by Cen~A in an earlier phase where it was potentially 200 times more luminous \citep{bell+22,taylor+23}.
These considerations render Cen~A the most likely close-by UHECR-emitter with a contribution that lies around 10\%. However, the conclusion depends on extragalactic magnetic field models, and the possibility that Cen~A's contribution could even approach $\sim100$\% above the ankle has also been discussed~\citep{kim13,Eichmann:2017iyr,deOli+20,mollerach+24}.

UHE neutrinos enable us to probe Cen~A's contribution to the observed UHECR flux. The expected neutrino flux associated with UHECRs from Cen~A's jet is shown in Figure~\ref{fig:cenA-flux}, where the solid bands denote the jet neutrino flux for a more or less optimistic contribution of Cen~A to the UHECR flux, $(3-100)$\% at 40~EeV based on the injection slope $q$. 
The predicted fluxes are lower than the sensitivity predictions of future and current experiments POEMMA \citep{POEMMA20}, GRAND \citep{GRAND20}, KM3Net \citep{KM3NET24}, and IceCube-Gen2 \citep{ICECUBE-GEN2,ICECUBE-ICRC21}. 
However, we could test the case with Cen~A's 100\% contribution to the UHECR flux.

Beyond the jet, we compare contributions from Cen A’s giant lobes with observatory resolutions, making it essential to account for the angular resolution of each observatory. The giant lobes of Cen~A have a significant angular extent of $\theta_{\rm lobe} \sim 4^\circ$. When this size exceeds the observatory’s point spread function (PSF), i.e., $\theta_{\rm lobe} > \theta_{\rm psf}$, the sensitivity decreases by a factor of $\sim \theta_{\rm lobe}/\theta_{\rm psf}$.
The PSF varies across observatories, with resolutions either at the degree or sub-degree level: POEMMA’s $\theta_{\rm psf} \lesssim 1.5^\circ$ \citep{POEMMA20}, IceCube Gen2’s $\theta_{\rm psf} \lesssim 1^\circ$ \citep{ICECUBE23res}, KM3NeT’s $\theta_{\rm psf} \sim 0.1^\circ$ for tracks and $1^\circ$ for showers \citep{KM3NET24b}, and GRAND’s $\theta_{\rm psf} \lesssim 1^\circ$ \citep{kotera24res}.
In contrast, these resolution effects are negligible for detecting emissions from the jet, which has an angular size of at most $0.1^\circ$, corresponding to its physical size of 1–10 kpc. Finally, we include considerations for neutrinos associated with delayed propagation (discussed in detail below).

%%%%%%%%%%%%%
%%%%%%%%%%%%%
%%%%%%%%%%%%%

\subsection{Neutrino flux due to UHECR Propagation from Cen~A to Earth}
The UHECR composition is heavy with CNO-like or slightly heavier UHECRs dominating the highest energies \citep{AUGER23}.
We therefore calculate the flux of neutrinos $\phi_\nu$ produced during the propagation of CNO-like UHECRs from Cen~A to Earth. 
For such $p\gamma$ and $A\gamma$ interactions, we consider the cosmic microwave background (CMB) and extragalactic background light (EBL) \citep{dole+06} as photon targets. The actual propagation time may exceed the ballistic time depending on the strength and coherence length of the extragalactic magnetic fields, e.g., \citep[][]{globus+17a};
although the actual intergalactic magnetic field may be $\sim0.1-10$nG. e.g., \citep{vazza+17b}. Fields as large as $B \sim 0.1~\mu$G have also been considered within the Council of Giants \citep{vallee+02,taylor+23}. 
We can express the propagation time $t_{\rm prop}$ from the source to Earth in the diffusive regime for a particle with Larmor radius $\mathcal{R} \ll \ell_c$, such that $\ell_c$ is the coherence length of the magnetic field; then $t_{\rm prop} \sim d_{\rm cenA}^2 \ell_c^{-2/3}\mathcal{R}^{-1/3}/c$. 
For a UHE CNO particle with $E = 5 \times 10^{19}$~eV, $t_{\rm prop} \sim 40$~Myr for $\ell_c = d_{\rm cenA}$\footnote{$d_{\rm cenA}$ is the distance to Cen~A, where $d_{\rm cenA}\simeq3.5$~Mpc, e.g., \citep[][]{ferrarese+07,majaess10}.}. This diffusive method maximizes $t_{\rm prop}$, but should not exceed the expected jet age, as these considerations assume that UHECRs were accelerated in the jet in a distant past and were delayed during their propagation. 
Cen~A's jet inferred age from the synchrotron age measurements of the lobes is $\approx 30$~Myr \citep{hardcastle+09}, and hence, we can consider the maximum propagation time to be $t_{\rm prop} \approx 30$~Myr.

%which is consistent with the above limit.
%The propagation time can be maximized
%We can then maximize the propagation time to Earth by setting it to the expected jet age, $t_{\rm prop} \approx 30$~Myr, which is inferred from the synchrotron age measurements of the lobes \citep{hardcastle+09}. 

\begin{figure}
	\begin{centering}
\includegraphics[width=0.48\textwidth,clip=false,trim= 0 0 0 0]{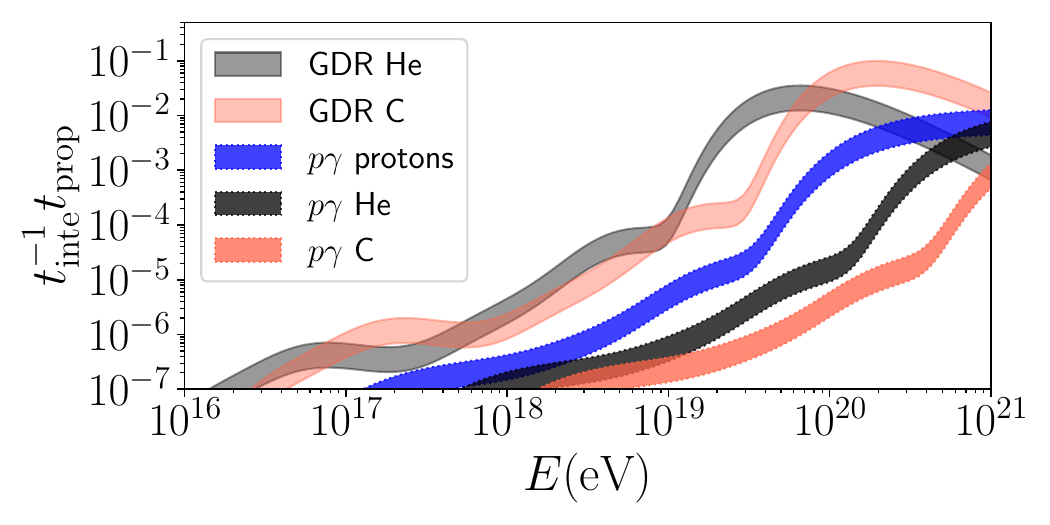}
		\caption{Expected neutrino production efficiency during propagation from Cen~A parameterized by $ t_{\rm inte}^{-1} t_{\rm prop}$, for interactions due to GDR and $p\gamma$ for protons, Helium (He), and carbon (C). The width of the plots determines the effect of minimum and maximum delays incurred during UHECR propagation due to magnetic fields.
		}
		\label{fig:cenA-prop}
	\end{centering}
\end{figure}

We can then express the neutrino flux due to interactions during propagation from the source~\citep{murase+10},
\begin{equation}\label{eq:nu-eq}
	E_\nu^2 \phi_\nu (E_\nu) \approx
	\begin{cases}
		 \frac{3}{2} \alpha_{p\gamma} t^{-1}_{p\gamma-\rm inte} t_{\rm prop} E^2 \phi_{\rm UHECR},& \text{for }p\gamma  \\
		\frac{1}{2} \alpha_{\rm GDR} t^{-1}_{A\gamma-\rm inte} t_{\rm prop} E^2 \phi_{\rm UHECR},& \text{for }A\gamma  
	\end{cases}
\end{equation}
where $t_{\rm inte}$ is the interaction time for different processes, $E$ is the parent particle energy, and $E_\nu = \alpha E$ the neutrino energy, with $\alpha_{p\gamma} \approx 1/(20)$ for $p\gamma$ interactions, and $\alpha_{\rm GDR} \approx 1/(2000 A)$ for $A\gamma$ interactions \footnote{The factor of 3/2 in photomeson interactions reflects the charged pion to the neutral pion ratio and the number of neutrinos, while the factor of 1/2 in GDR interactions reflects the fact that only neutrons result in neutrino emission.}.
Figure~\ref{fig:cenA-prop} shows the expected optical depth $t^{-1}_{\rm inte} t_{\rm prop}$ for a CMB+EBL photon field.
The width of the resulting optical depth denotes the extent of the effects of delays due to magnetic fields, where the lower limits are for ballistic propagation, and upper limits for the maximum expected delays for propagation from Cen~A.

Using Eq.~\ref{eq:nu-eq}, we find that for the UHECR flux reported in the direction of Cen~A \citep{AUGER22}, the expected neutrino flux due to this maximized propagation time is \emph{i)} $E_\nu^2 \phi_\nu \sim 5 \times 10^{-16}$~erg cm$^{-2}$ s$^{-1}$ at $E_\nu \sim 10^{18}$ eV due to $p\gamma$ interactions, and \emph{ii)} $E_\nu^2 \phi_\nu \sim 3 \times 10^{-16}$~erg cm$^{-2}$ s$^{-1}$ at $E_\nu \sim 10^{15}$ eV due to $A\gamma$ interactions.
These values are either comparable or lower than the expected source neutrino flux.
We add these considerations to Figure~\ref{fig:cenA-flux} in light brown for reference.
%\dam{Need to comment on the assumed propagation time; if lower, the nu flux is lower...}

%%%%%%%%%%%%%%
%%%%%%%%%%%%%%
%%%%%%%%%%%%%%
%%%%%%%%%%%%%%

\subsection{Neutrino Flux due to propagation in Cen~A's Giant Lobes}
Cen~A exhibits two giant lobes beyond its jets that reach distances of order of $S_{\rm gl}\simeq 280$kpc \citep{feain+11}, where $S_{\rm gl}$ is the size of the giant lobes.
Importantly, radio data from \citet{hardcastle+09}, along with the gamma-ray detection in the lobes puts the magnetic field at the level of $B_{\rm gl}\sim \mu$G \citep{FERMI10}.
Hence, if UHECRs are injected in the giant lobes after acceleration in the jet, their diffusion time is $t_{\rm diff} \approx 3 S_{\rm gl}^2/(\mathcal{R} c)$.
$t_{\rm diff}$ can be as large as 300Myr, in the case of Helium at $E > 10^{19}$~eV. 
However, the synchrotron age of Cen~A's giant lobes is $t_{\rm gl} \lesssim 30$~Myr \citep{hardcastle+09}, and therefore $t_{\rm diff}$ is bound by $\lesssim 30$~Myr.
%Considering that UHECRs should escape the jet quite isotropically \cite{mbarek+19}, we assume that a similar UHECR power is injected in the lobes as is detected on Earth. \dam{what do you mean?}
The CMB and EBL are the dominant radiation fields in the lobes considering the large distances from the base of the jet. We find that the Giant Lobes have a similar neutrino contribution to propagation from Cen~A as shown in Figure~\ref{fig:cenA-flux}.

The cosmogenic gamma-ray flux from Cen~A has been estimated to be $\sim{10}^{-15}-{10}^{-13}~{\rm erg}~{\rm cm}^{-2}~{\rm s}^{-1}$, depending on its contribution to the observed UHECR flux~\cite{Der+09,deOli+20}. In particular, Ref.~\cite{deOli+20} assumed the extreme case where almost all UHECRs come from Cen~A, whereas our prediction can be lower partly because of the lower contribution from Cen~A. 

%%%%%%%%%%%%%%
%%%%%%%%%%%%%%
%%%%%%%%%%%%%%
\subsection{Potential for neutrino detection}
\subsubsection{PeV to EeV neutrinos}
In this energy range, the Earth is opaque to neutrinos, and neutrino observatories are most sensitive to horizontal and downward-going neutrinos above the horizon.
Considering Cen~A's location in the sky, 
GRAND \citep[Fig.~8 in][]{GRAND20}, POEMMA \citep{POEMMA20}, and IceCube-Gen2 radio \citep{ICECUBE-GEN2} should be the most sensitive instruments. 
POEMMA's 3yr resolution is obtained with rescaling of the sensitivity presented in Fig.~2 in \citet{POEMMA20}, taking into account that the neutrino background rate is negligible at each observing run. 
As for IceCube-Gen2 radio, the flux sensitivity is converted from Fig.~3 in \citet{ICECUBE-ICRC21} for Cen~A's declination, $\delta\sim-45$~deg, by averaging the curves for $\delta=-40$~deg and $\delta=-50$~deg. This is consistent with another estimate of the sensitivity based on rescaling of the RNO-G sensitivity \citep{aguilar+21}, assuming that RNO-G has a similar design to Gen2 radio with a station ratio of $\sim 35/200$. We note that the GRAND and Gen2 radio sensitivities depend on the source declination, and both detectors are complementary for the purpose of detecting nearby point sources.   
We emphasize here that an increased sensitivity of these instruments, along with a larger-than-expected contribution from Cen~A to the overall UHECR flux are needed for a potential neutrino detection from Cen~A's jet. With more than a decade of observations by IceCube-Gen2 radio, we could test the extreme case where Cen~A contributes to $\sim100$\% of the UHECR flux.

%southern hemisphere observatories such as IceCube-Gen2 are the most promising \citep{martinez+16}. 
%Figure~\ref{fig:cenA-flux} shows the expected resolution of IceCube-Gen2, based on a rescaling of the RNO-G sensitivity \citep{icecube-p,aguilar+21}.
%We find that even if the radio observatory of IceCube-Gen2 is potentially the most sensitive to neutrinos from Cen~A's jet, a detection is unlikely in the near-future. 
%However, it is worth noting that a detection in this energy band would enable a more accurate and direct characterization of the UHECR injection slope compared to other methods.

%Indeed, even if the flux is comparable for both UHECR injection spectra, we consistently obtain more neutrinos for q = 2 at the $\sim$ PeV level, and for q=1 at the $\sim 0.1$~EeV level.

\subsubsection{TeV to PeV Neutrinos}

%\begin{figure}
%	\centering
 %\includegraphics[width=0.45\textwidth,clip=false, trim= 0 0 0 0]{plots/q-beta.png}
%	\caption{Expected events from Cen~A for different values of $\beta$ compared with IceCube's neutrino background within 1$^\circ$ of Cen~A. We can see that for the Giant Lobe/propagation contribution, the number of expected events deviates from the background for $\beta_{\rm Km3Net} \simeq 10$.}
%	\label{fig:data}
%\end{figure}

%In Figure~\ref{fig:data} we compare the expected neutrino flux from Figure~\ref{fig:cenA-flux} with the IceCube 10-yr background for events within $1^\circ$ of Cen~A for different UHECR injection slopes $q$, along with the expected yield from Cen~A's giant lobes and propagation from the source. 
%We then scale the flux by a factor $\beta_{\rm IC}$ to check against different neutrino detection resolutions. 
%In a nutshell, $\beta_{\rm IC}$ is a proxy for a potential improved resolution to the IceCube 10yr detection.
%For IceCube-Gen2, the corresponding $\beta$ is $\beta_{\rm IC} \simeq 10 \beta_{\rm Gen2}$ 
%since a factor of $\simeq$10 increase in resolution compared to IceCube is expected \citep{ICECUBE-GEN2-14}.

We can see from Figure~\ref{fig:cenA-flux} that with planned neutrino detectors, it is challenging to detect neutrinos at the PeV level from Cen~A's jet, even with the advent of KM3Net \citep{KM3NET24}.
KM3Net's resolution is obtained by averaging based on Cen~A's declination from Figure~B.6. in \citet{KM3NET24}.
%\citep{martinez+16}. 
A neutrino detection in the direction of Cen~A in the near-future could mean that \emph{i)} sources other than the jet are neutrino-emitters, such as radiatively inefficient accretion flows (RIAFs) and black hole coronae, e.g., \citep[][]{Mur+20,kheirandish+21,mbarek+24}, or \emph{ii)} that the increased photon density associated with a potential flare increases the neutrino yield, a possibility put forward to explain the association of UHE neutrinos with TXS-0506+056 \citep{TXS065-18}. 
A much longer acceleration time and/or confinement of UHECRs in the vicinity of Cen~A is unlikely, because it would lead to a complete depletion of heavy ions and thus contradict the high atomic mass of detected UHECRs \citep{mbarek+23}.
%considering that for $q=2$  $\beta_{\rm IC} > 10^3$ and $\beta_{\rm Gen2} > 10^2$ for the signal to deviate from the background. A deviation from the background is necessary to claim a potential association with Cen~A.
%The advent of KM3Net \citep{martinez+16}, which is $\sim 50 \times$ more sensitive to sources located in the southern hemisphere in the TeV-PeV range (compared to IceCube) could be helpful in this sense.

%%%%%%%%%%%%%%%%%%%%%
%%%%%%%%%%%%%%%%%%%%%
%%%%%%%%%%%%%%%%%%%%%

\section{Conclusions}
In this paper, we calculated the UHE neutrino flux from the Cen~A jet expected in the espresso-shear acceleration mechanism. We used the method outlined in \citep{mbarek+23} to 
propagate seed CRs in a simulated relativistic jet with realistic parameters (size, aspect ratio, Lorentz factor), augmented with appropriate photon fields from the inner core of the jet \citep{zhang+23}. 
The main conclusion of the paper is that we need a quite large contribution from Cen~A to the UHECR flux in order to have a neutrino detection at the EeV level in the near-future. In other words, deeper neutrino observations reaching $10^{-13}-10^{-12}~{\rm erg}~{\rm cm}^{-2}~{\rm s}^{-1}$ can be used for constraining Cen~A's contribution to the UHECR flux independent of uncertain magnetic fields. 

Several remarks can be summarized as follows:
\begin{itemize}
    \item We set constraints on the expected neutrino flux from the Cen~A jet from 10 TeV to more than a few EeV, and find that neutrinos with $E_\nu > 10^{17}$eV from Cen~A's jet have a low flux compared to the resolutions of current and planned missions. 
    \item A detection of a higher flux would imply the optical depth of the jet/cocoon/giant lobes to be large enough to deplete heavy ions; alternatively, a burst of neutrinos may come from an unusual enhancement of the photon background, e.g., a flare.    
    %Increasing the probability of UHECR interactions contradicts observations either \emph{i)} of likely jet properties (magnetic fields, jet lobes), or \emph{ii)} of the heavy composition of UHECRs.
    %\item The best perspective for detection is represented by PeV neutrinos produced by the $\beta-$decay of neutrons from the photodisintegration of UHECRs with the KM3Net detector in the northern hemisphere.
    
    \item A TeV--PeV neutrino detection in the direction of Cen~A in the near future would likely not be associated with UHECR production, but could come from other regions in the AGN such as the disk corona, e.g., \citep[][]{Mur+20,kheirandish+21,mbarek+24}, ultrafast outflows, e.g., \citep[][]{peretti+23}, or neighboring sources to Cen~A. Although the neutrino flux from the corona has been calculated~\cite{kheirandish+21}, given that the accretion rate is estimated to be very low for FR-I jets, Cen~A is less likely to have a standard accretion disk and the disk will be RIAFs. Cosmic-ray acceleration in RIAFs has been studied~\citep{kimura+15,Kim+19}, and the neutrino flux toward Cen~A can reach $\sim {\rm a~few}\times{10}^{-13}~{\rm erg}~{\rm cm}^{-2}~{\rm s}^{-1}$ in this scenario~(see Fig.~5 of \citep{Fuj+15}).
\end{itemize}

\section{Acknowledgements}
We would like to thank Mauricio Bustamante for carefully reading our manuscript on behalf of the GRAND Collaboration. We also thank Brian Allen Clark, Kumiko Kotera, and Michael Larson for helpful conversations. We also thank B. Theodore Zhang for providing the spectral template for Cen~A's jet. 
Simulations were performed on computational resources provided by the University of Chicago Research Computing Center. 
The work of K.M. is supported by the NSF grant Nos.~AST-2108466, AST-2108467, and AST-2308021, and the JSPS KAKENHI grant Nos.~20H01901 and 20H05852 (KM).
D.C. was partially supported by NSF through grant AST-2308021.

\bibliography{Total}
%\bibliographystyle{aasjournal.bst}
%\bibliography{apssamp}

\end{document}